\documentclass[a4paper,11pt]{article}
\usepackage{pos}
\usepackage[compat=1.1.0]{tikz-feynman}
\usepackage{xcolor}
\usepackage{graphicx,epsfig,amsmath}
\pgfmathsetmacro\sizedot{1.1}
\pgfmathsetmacro\sizesqdot{2.0}
\pgfmathsetmacro\sizerectangle{0.8}
\pgfmathsetmacro\sizecirc{0.55}
\pgfmathsetmacro\sizecrodot{1.0}

\newcommand{\lentrichromo}{28}
\newcommand{\lenboxchromo}{40}

\newcommand{\lentrismall}{20}

\newcommand{\lentriftop}{28}
\newcommand{\lenboxftop}{40}

\newcommand{\lentrieq}{20}
\newcommand{\lenboxeq}{28}

\newcommand{\mhh}{m_{hh}}
\newcommand{\msbar}{\ensuremath{\overline{\mathrm{MS}}}}

\def\eps{\epsilon}

\title{Subleading operators and $\gamma_5$-scheme dependence in SMEFT
  for Higgs boson pair production}
\ShortTitle{$\gamma_5$-scheme dependence in SMEFT for Higgs pair production}

\author[a]{Stefano Di Noi}
\author[a]{Ramona Gr\"ober}
\author*[b]{Gudrun Heinrich}
\author[b]{Jannis Lang}
\author[c]{Ludovic Scyboz}
\author[d,e]{Marco Vitti}

\affiliation[a]{
Dipartimento di Fisica e Astronomia ``G. Galilei", Universit\`a di Padova, Italy, \\ and Instituto
Nazionale di Fisica Nucleare, Sezione di Padova, 35131 Padova, Italy}

\affiliation[b]{Institute for Theoretical Physics, Karlsruhe Institute of Technology (KIT), \\Wolfgang-Gaede-Str.~1, 76131 Karlsruhe, Germany}

\affiliation[c]{School of Physics and Astronomy, Monash University, Clayton VIC 3800, Australia}

\affiliation[d]{%
  Institute for Theoretical Particle Physics, Karlsruhe Institute of Technology (KIT), \\
  Wolfgang-Gaede-Str.~1, 76131 Karlsruhe, Germany
}
\affiliation[e]{%
  Institute for Astroparticle Physics, Karlsruhe Institute of Technology (KIT), \\
  Hermann-von-Helmholtz-Platz 1, 76344 Eggenstein-Leopoldshafen, Germany
}

\emailAdd{gudrun.heinrich@kit.edu}
\emailAdd{jannis.lang@kit.edu}
\emailAdd{stefano.dinoi@phd.unipd.it}
\emailAdd{ramona.groeber@pd.infn.it}
\emailAdd{ludovic.scyboz@monash.edu}
\emailAdd{marco.vitti@kit.edu}

\abstract{The calculation of contributions from the chromomagnetic and
  four-top-quark-operators within Standard Model Effective Field
  Theory (SMEFT) to Higgs boson pair production in gluon fusion is
  presented, in combination with NLO QCD corrections. Here we focus on the $\gamma_5$-scheme dependence introduced by the four-top-quark-operators and the interplay with other operators contributing to this process in SMEFT.}

\FullConference{Loops and Legs in Quantum Field Theory (LL2024)\\
 14-19, April, 2024\\
Wittenberg, Germany\\}

\begin{document}
\maketitle

\section{Introduction}

Higgs boson pair production plays a special role in the LHC program
since it is the prime process to constrain the trilinear Higgs
coupling. The gluon fusion production mode has the largest cross
section, therefore a lot of effort has been put into providing increasingly accurate predictions for this process.

In these proceedings we focus on the description of $gg\to hh$ within Standard Model Effective Field Theory (SMEFT), combining leading and subleading operators with NLO QCD corrections in the Standard Model (SM) as described in more detail in Refs.~\cite{Heinrich:2023rsd,Lang:2024bsk}.
The contributions to $gg\to hh$ of the leading operators in SMEFT have been calculated in Ref.~\cite{Heinrich:2022idm} and have been implemented in the  {\tt Powheg-Box-V2} event generator~\cite{Alioli:2010xd}, while the calculation of the operators in the HEFT framework has been presented in Refs.~\cite{Buchalla:2018yce,Heinrich:2020ckp}, including the NLO QCD corrections calculated in Ref.~\cite{Borowka:2016ehy}.
The chromomagnetic and the 4-top-operators are suppressed by loop
factors compared to the leading operators when the potential UV completion is assumed to be a weakly coupling and renormalisable quantum field theory~\cite{Arzt:1994gp,Buchalla:2022vjp}.
We will demonstrate that these operators are intricately related
through a $\gamma_5$-scheme dependence; the scheme dependence only cancels when they are consistently combined in a renormalised amplitude, as has been shown in Ref.~\cite{DiNoi:2023ygk} for the case of single Higgs production and in Ref.~\cite{Heinrich:2023rsd} for double Higgs production.

\section{Operators contributing to $gg\to hh$ beyond the leading order}

Any bottom-up EFT  is defined by its degrees of freedom, 
its  symmetries and a power counting scheme.
SMEFT~\cite{Buchmuller:1985jz,Grzadkowski:2010es,Brivio:2017vri,Isidori:2023pyp} builds on the field content and gauge symmetries of the SM and its main power counting, 
which relies on the counting of the canonical (mass) dimension, expanding in inverse powers of a new physics scale $\Lambda$ which suppresses operators beyond dimension-4.
The dominant contributions are expected to be described by dimension-6 operators, on which we focus here.
We also impose a flavour symmetry $U(2)_q\times U(2)_u\times U(3)_d$ in the quark sector, which forbids chirality flipping operators bilinear in light quarks (including $b$-quarks), such that only 4-top-operators remain.
Further, we neglect operators whose contributions involve diagrams with electroweak particles propagating in the loop.

With these restrictions, the dimension-6 CP even operators that contribute to $gg\to hh$, after electroweak symmetry breaking and in the unitary gauge, are given by
\newpage
\begin{equation}
\begin{aligned}
    {\cal L}_{\text{SMEFT}}&\supset
-\left(\frac{m_t}{v}\left(1+v^2\frac{C_{H,\text{kin}}}{\Lambda^2}\right)-\frac{v^2}{\sqrt{2}}\frac{C_{tH}}{\Lambda^2}\right)\,h\,\bar{t}\,t
-\left(m_t \frac{C_{H,\text{kin}}}{\Lambda^2}-\frac{3v}{2\sqrt{2}}\frac{C_{tH}}{\Lambda^2}\right)\,h^2\,\bar{t}\,t
\\
&-\left(\frac{m_h^2}{2v}\left(1+3v^2\frac{C_{H,\text{kin}}}{\Lambda^2}\right)-v^3\frac{C_H}{\Lambda^2}\right)\,  h^3
+\frac{C_{HG}}{\Lambda^2}\left(v\,h+\frac{1}{2}\,h^2\right)\, G^a_{\mu \nu} G^{a,\mu \nu}
\\
&+g_s\bar{t}\,\gamma^{\mu}T^a\,t\,G_{\mu}^a
+\frac{C_{tG}}{\Lambda^2}\sqrt{2}\left(h+v\right)\left(\bar{t}\,\sigma^{\mu\nu}T^a\,t\,G_{\mu\nu}^a\right)
\\ &
+\frac{C_{Qt}^{(1)}}{\Lambda^2}\bar{t}_L\gamma^\mu t_L\bar{t}_R\gamma_\mu t_R
+\frac{C_{Qt}^{(8)}}{\Lambda^2}\bar{t}_L\gamma^\mu T^at_L\bar{t}_R\gamma_\mu T^a t_R
\\ &
+\frac{C_{QQ}^{(1)}}{\Lambda^2}\bar{t}_L\gamma^\mu t_L\bar{t}_L\gamma_\mu t_L
+\frac{C_{QQ}^{(8)}}{\Lambda^2}\bar{t}_L\gamma^\mu T^at_L\bar{t}_L\gamma_\mu T^a t_L
\\ &
+\frac{C_{tt}}{\Lambda^2}\bar{t}_R\gamma^\mu t_R\bar{t}_R\gamma_\mu t_R\;,\label{eq:LagSMEFT_expanded}
\end{aligned}
\end{equation}
where  $\sigma^{\mu \nu}=\frac{i}{2}\left[\gamma^{\mu},\gamma^{\nu}\right]$ and $\tilde{\phi}=i\sigma_2\phi$ is the charge conjugate of the Higgs doublet.
The first two lines in Eq.~\eqref{eq:LagSMEFT_expanded} contain the leading EFT operators, which have been studied in Ref.~\cite{Heinrich:2022idm}. The remaining lines contain the chromomagnetic operator and the 4-top operators, where we use ${\cal O}_{QQ}^{(1),(8)}$, related to the corresponding operators in the  Warsaw basis~\cite{Grzadkowski:2010es} by
%
\begin{equation}
\begin{aligned}
    C_{QQ}^{(1)}&=C_{qq,\,\text{Warsaw}}^{(1)\,3333}-\frac{1}{3}C_{qq,\,\text{Warsaw}}^{(3)\,3333}\;\; , \;\;
    C_{QQ}^{(8)}=4\,C_{qq,\,\text{Warsaw}}^{(3)\,3333}\;.
\end{aligned}
\end{equation}
We emphasize that $v$ denotes the full vacuum expectation value including 
a higher dimensional contribution of $C_H/\Lambda^2$ and the relation between  the top-Yukawa parameter $y_t$ of the SM Lagrangian and the top quark mass is given by
\begin{equation}
m_t=\frac{v}{\sqrt{2}}\left(y_t -\frac{v^2}{2}\frac{C_{tH}}{\Lambda^2}\right)\;.
\end{equation}


In the following, we will explain the notions of `leading' and `subleading' operators we have used above.
In SMEFT, the operators are ordered by their canonical dimension, i.e.\ the
expansion relies on powers in $E/\Lambda$.
However, in a perturbative expansion, in particular in the combination
of an EFT expansion with expansions in a SM coupling, loop suppression
factors also play a role. Therefore, a classification of operators
into {\em potentially} tree-level induced  and 
loop-generated operators~\cite{Arzt:1994gp,Isidori:2023pyp} can be a powerful criterion to identify the relative importance of dimension-6 operators in SMEFT. Loop-generated operators carry an implicit loop factor ${\mathbf L}=\left(16\pi^2\right)^{-1}$, they are typically given by operators containing at least one field strength tensor.
We use a boldface notation for the loop factors that are not SM-induced.
 The loop factors can be derived by
 supplementing the SMEFT expansion by a chiral counting of operators~\cite{Buchalla:2022vjp}, see also~\cite{Guedes:2023azv}.
 Such a classification cannot be derived without making some minimal UV assumptions, which are however
quite generic, assuming renormalisability and weak coupling of the underlying UV complete theory.
Under these assumptions, and if the Wilson coefficients $C_i$ in the SMEFT expansion are considered to be of similar magnitude, it makes sense to expand in
\begin{equation}
  \frac{C_i}{\Lambda^a} \, \times \, 1/(16\pi^2)^b\;.
  \end{equation}
  Fixing $a=2$ (dimension-6 operators), we call the operator contributions with $b=0$ `leading' and those with $b>0$ `subleading'.
  The above factors are to be combined with {\em explicit} loop factors $ L=1/(16\pi^2)^c$ from the SM perturbative expansion.
  We will see below that this classification is corroborated by observations from
renormalisation and the cancellation of scheme-dependent
terms~\cite{DiNoi:2023ygk}. 
Applying those rules to the Born contributions
and associating loop factors of QCD origin with  powers of $g_s$ leads to 
 ${\cal M}_\text{Born}\sim {\cal O}\left((g_s^2L)\Lambda^{-2}\right)$.
 
\subsection{Chromomagnetic operator insertions}
\label{sec:MEchromo}
The contribution of the chromomagnetic operator to the amplitude leads to the diagram types shown in Fig.~\ref{fig:feyndiag_chromo}. 
\begin{figure}[h]
\begin{center}
\begin{minipage}[c]{0.24\textwidth}
    \centering
\begin{tikzpicture} 
    \begin{feynman}[small]
        \vertex  (g1)  {$g$};
        \vertex  (gtt1) [square dot, scale=\sizesqdot,right=\lentrichromo pt of g1,color=gray] {};
        \vertex (htt) [dot,scale=\sizedot,below right =\lentrichromo pt of gtt1]  {};
        \vertex  (gtt2) [dot,scale=\sizedot,below left =\lentrichromo pt of htt] {};
        \vertex  (g2) [left=\lentrichromo pt of gtt2]  {$g$};
        \vertex (hhh) [dot,scale=\sizedot,right =\lentrichromo pt of htt]  {};
        \vertex (h1) [above right =\lentrichromo pt of hhh] {$h$};
        \vertex (h2) [below right =\lentrichromo pt of hhh] {$h$};
        \diagram* {
            (g1)  -- [gluon] (gtt1),
            (g2) -- [gluon] (gtt2),
            (h1)  -- [scalar] (hhh) -- [scalar] (h2),
            (htt) -- [scalar] (hhh),
            (gtt1) -- [fermion, line width = 1.5 pt] (htt) 
            -- [fermion, line width = 1.5 pt] (gtt2)
             -- [fermion, line width = 1.5 pt] (gtt1), 
            
        };
    \end{feynman}
\end{tikzpicture}
    \caption*{(a)}
\end{minipage} 
\begin{minipage}[c]{0.24\textwidth}
    \centering
\begin{tikzpicture} 
    \begin{feynman}[small]
        \vertex  (g1)  {$g$};
        \vertex  (gtt1) [square dot, scale=\sizesqdot,right= of g1,color=gray] {};
         \vertex (htt) [dot,scale=0.01,below right =\lentrichromo pt of gtt1,color=white]  {};
        \vertex  (gtt2) 
        [dot,scale=\sizedot,below left =\lentrichromo pt of htt] {};
        \vertex  (g2) [left=of gtt2]  {$g$};
        \vertex (hhh) [dot,scale=\sizedot,right =7 pt of htt] {};
        \vertex (h1) [above right =\lentrichromo pt of hhh] {$h$};
        \vertex (h2) [below right =\lentrichromo pt of hhh] {$h$};
        \diagram* {
            (g1)  -- [gluon] (gtt1),
            (g2) -- [gluon] (gtt2),
            (hhh)  -- [scalar] (gtt1),
            (h1) -- [scalar] (hhh) -- [scalar] (h2),
            (gtt1) -- [fermion, line width = 1.5 pt,quarter right] (gtt2)
             -- [fermion, line width = 1.5 pt, quarter right] (gtt1), 
            
        };
    \end{feynman}
\end{tikzpicture}
    \caption*{(b)}
\end{minipage}
\begin{minipage}[c]{0.24\textwidth}
    \centering
\begin{tikzpicture} 
    \begin{feynman}[small]
        \vertex  (g1)  {$g$};
        \vertex  (gtt1) [square dot,scale=\sizesqdot,right=\lentrichromo pt of g1,color=gray] {};
        \vertex  (htt1) [dot,scale=\sizedot,right=\lenboxchromo pt of gtt1]  {};
        \vertex  (gtt2) [dot,scale=\sizedot,below=\lenboxchromo pt of gtt1] {};
        \vertex  (htt2) [dot,scale=\sizedot,below=\lenboxchromo pt of htt1] {};
        \vertex  (g2) [left=\lentrichromo pt of gtt2]  {$g$};
        \vertex (h1) [right =\lentrichromo pt of htt1] {$h$};
        \vertex (h2) [right =\lentrichromo pt of htt2] {$h$};
        \diagram* {
            (g1)  -- [gluon] (gtt1),
            (g2) -- [gluon] (gtt2),
            (h2)  -- [scalar] (htt2),
            (h1)  -- [scalar] (htt1),
            (gtt1) -- [fermion, line width = 1.5 pt] (htt1)  -- [fermion, line width = 1.5 pt] (htt2)
            -- [fermion, line width = 1.5 pt] (gtt2)
             -- [fermion, line width = 1.5 pt] (gtt1), 
        };
    \end{feynman}
\end{tikzpicture}
    \caption*{(c)}
\end{minipage}
\begin{minipage}[c]{0.24\textwidth}
    \centering
\begin{tikzpicture} 
    \begin{feynman}[small]
        \vertex  (g1)  {$g$};
        \vertex  (ghtt1) [square dot, scale=\sizesqdot,right=\lentrichromo pt of g1,color=gray] {};
        \vertex (htt) [dot,scale=\sizedot,below right =\lentrichromo pt of gtt1]  {};
        \vertex  (gtt2) [dot,scale=\sizedot,below left =\lentrichromo pt of htt] {};
        \vertex  (g2) [left=\lentrichromo pt of gtt2]  {$g$};
        \vertex (h1) [right =\lenboxchromo pt of ghtt1] {$h$};
        \vertex (h2) [below right =\lentrichromo pt of htt] {$h$};
        \diagram* {
            (g1)  -- [gluon] (ghtt1),
            (g2) -- [gluon] (gtt2),
            (h1)  -- [scalar] (ghtt1),
            (h2) -- [scalar] (htt),
            (ghtt1) -- [fermion, line width = 1.5 pt] (htt) 
            -- [fermion, line width = 1.5 pt] (gtt2)
             -- [fermion, line width = 1.5 pt] (ghtt1), 
        };
    \end{feynman}
\end{tikzpicture}
    \caption*{(d)}
\end{minipage}
    \caption{\label{fig:feyndiag_chromo} Feynman diagrams involving insertions of the chromomagnetic operator. 
    The gray squares denote insertions of the chromomagnetic operator.}
\end{center}
\end{figure}
At first sight, the diagrams are at one-loop order.
However, taking into account that the chromomagnetic operator belongs to the class of operators that, in 
renormalisable UV completions, can only be generated at loop level,  the order in the power counting is  ${\cal M}_{tG}\sim {\cal O}\left((g_s^2L){\mathbf L}\,\Lambda^{-2}\right)$, 
which contains an additional factor ${\mathbf L}=1/(16\pi^2)$ relative  to the leading Born diagrams.

The diagrams of type (a), (b) and (d) are UV divergent even though they constitute the leading order contribution of $C_{tG}$ to the gluon fusion process. 
This behaviour is well known~\cite{Degrande:2012gr,Maltoni:2016yxb,Deutschmann:2017qum} and leads to a renormalisation of 
$C_{HG}^0=\mu^{2\eps}\left(C_{HG}+\delta_{C_{HG}}^{C_{i}}\right)$,  
which in the $\msbar$ scheme takes the form~\cite{Alonso:2013hga,Maltoni:2016yxb,Deutschmann:2017qum}
\begin{equation}
    \delta _{C_{HG}}^{C_{tG}}=\frac{\left(4\pi e^{-\gamma_E}\right)^\eps}{16\pi^2\eps}\frac{4\sqrt{2}g_s m_t}{v}T_F\,C_{tG}\;.
\end{equation}
%

\subsection{Amplitude structure involving four-top operators}
\label{sec:ME4t}

Four-top operators appear first at two-loop order in  Higgs- or di-Higgs production in gluon-fusion, where the two loops are explicit.
Taking into account the loop-generated nature of the  chromomagnetic operator, their contribution is of the same order in the power counting as the chromomagnetic operator,
i.e.\ ${\cal M}_\text{4-top}\sim {\cal O}\left((g_s^2L){\mathbf L}\,\Lambda^{-2}\right)$.
The complete set of diagrams involving 4-top-operators in $gg\to hh$ can be found in Ref.~\cite{Heinrich:2023rsd}, here we only show in Fig.~\ref{fig:feyndiag_4top} those where a contraction of a one-loop subdiagram  leads
 to topologies of Fig.~\ref{fig:feyndiag_chromo}. 

\begin{figure}[h]
\begin{center}
\begin{tabular}{ c  c  c }

\begin{minipage}[c]{0.3\textwidth}
    \centering
\begin{tikzpicture} 
    \begin{feynman}[small]
        \vertex  (g1)  {$g$};
        \vertex  (gtt1) [dot,scale=\sizedot,right =\lentrismall pt  of g1] {};
        \vertex (4F) [dot, scale = \sizesqdot,color=gray,right=\lentriftop/2 pt of gtt1] {};
        \vertex (htt) [dot,scale=\sizedot,below right =\lentriftop pt of 4F] {};
        \vertex  (gtt2) [dot,scale=\sizedot,below left =\lentriftop pt of htt] {};
        \vertex  (g2) [left=\lentrismall+\lentriftop/2 pt of gtt2]  {$g$};
        \vertex (hhh) [dot,scale=\sizedot,right =\lentriftop pt  of htt] {};
        \vertex (h1) [above right =\lentriftop pt of hhh] {$h$};
        \vertex (h2) [below right =\lentriftop pt of hhh] {$h$};
        \diagram* {
            (g1)  -- [gluon] (gtt1),
            (g2) -- [gluon] (gtt2),
            (hhh)  -- [scalar] (htt),
            (h1)  -- [scalar] (hhh) -- [scalar] (h2),
            (4F) -- [fermion, line width = 1.5 pt] (htt) 
            -- [fermion, line width = 1.5 pt] (gtt2) -- [fermion, line width = 1.5 pt] (4F), 
             (4F) -- [fermion, line width = 1.5 pt, half left] (gtt1) -- [fermion, line width = 1.5 pt, half left] (4F)
};
    \end{feynman}
\end{tikzpicture}
    \caption*{(a)}
\end{minipage}
&
\begin{minipage}[c]{0.3\textwidth}
    \centering
\begin{tikzpicture} 
    \begin{feynman}[small]
        \vertex  (g1)  {$g$};
        \vertex  (gtt1) [dot,scale=\sizedot,right =\lentrismall pt  of g1] {};
        \vertex (4F) [dot, scale = \sizesqdot,color=gray,right=\lentriftop/2 pt of gtt1] {};
        \vertex (htt1) [dot,scale=\sizedot,right =\lenboxftop pt of 4F] {};
        \vertex (htt2) [dot,scale=\sizedot,below =\lenboxftop pt of htt1] {};
        \vertex  (gtt2) [dot,scale=\sizedot,left =\lenboxftop pt of htt2] {};
        \vertex  (g2) [left=\lentrismall+\lentriftop/2 pt of gtt2]  {$g$};
        \vertex (h1) [right =\lentrismall+\lentriftop/2  pt of htt1] {$h$};
        \vertex (h2) [right =\lentrismall+\lentriftop/2  pt of htt2] {$h$};
        \diagram* {
            (g1)  -- [gluon] (gtt1),
            (g2) -- [gluon] (gtt2),
            (h1)  -- [scalar] (htt1),
            (htt2) -- [scalar] (h2),
            (4F) -- [fermion, line width = 1.5 pt] (htt1) -- [fermion, line width = 1.5 pt] (htt2) 
            -- [fermion, line width = 1.5 pt] (gtt2) -- [fermion, line width = 1.5 pt] (4F), 
             (4F) -- [fermion, line width = 1.5 pt, half left] (gtt1) -- [fermion, line width = 1.5 pt, half left] (4F)
};
    \end{feynman}
\end{tikzpicture}
    \caption*{(b)}
\end{minipage}
\end{tabular}
    \caption{\label{fig:feyndiag_4top} Selected Feynman diagrams involving insertions of 4-top operators.
    The gray dots denote insertions of 4-top operators.}
\end{center}
\end{figure}
In the following we sketch the relation between those classes of diagrams,
focusing on  the $\gamma_5$-scheme dependence, which first has been investigated in this context in Ref.~\cite{DiNoi:2023ygk}.
The four-top operators contain chiral projection operators  $(\mathbb{I}\pm\gamma_5)/2$. It is well-known that the treatment of $\gamma_5$ in dimensional regularisation is highly non-trivial, as $\gamma_5$ is an intrinsically four-dimensional object, see e.g. Refs.~\cite{Jegerlehner:2000dz,Belusca-Maito:2023wah,Aebischer:2023nnv,Stockinger:2023ndm}.
We will consider two different schemes  for the continuation of $\gamma_5$ to $D=4-2\epsilon$ dimensions: na\"ive dimensional regularisation (NDR)~\cite{Chanowitz:1979zu} and the Breitenlohner-Maison-t'Hooft-Veltman (BMHV)~\cite{tHooft:1972tcz,Breitenlohner:1977hr} scheme. 

In our calculation, the treatment of $\gamma_5$ in the two schemes differs only by ${\cal O}(\eps)$ parts of the Dirac algebra in $D$ dimensions. Therefore, 
the renormalised result in the limit $D\to\,4$  differs between the two schemes only by terms
stemming from the $\eps$-dependent parts of the
Dirac algebra multiplying a pole of a loop integral.

The contributions to the gauge interactions from the diagrams in Fig.~\ref{fig:feyndiag_4top}
for the case of an on-shell external gluon evaluate to
\begin{equation}
    \begin{tikzpicture}[baseline=(4F)]
        \begin{feynman}[small]
            \vertex  (g1)  {$g$};
             \vertex (gtt1) [dot, scale=\sizedot, right= 30 pt of g1] {};
            \vertex  (4F) [dot,scale=\sizesqdot,right = 20 pt of gtt1, color = gray] {};
            \vertex  (t1) [above right= 25 pt of 4F] {$t$};
            \vertex (t2) [below right= 25 pt of 4F] {$t$};

            \diagram* {
                (g1)  -- [gluon] (gtt1),  
                (gtt1) -- [anti fermion, line width = 1.5 pt, half right] (4F) 
                -- [anti fermion, line width = 1.5 pt,half right] (gtt1),
                (t2) -- [fermion, line width = 1.5 pt]  (4F) -- [fermion, line width = 1.5 pt] (t1)
            };
        \end{feynman}
    \end{tikzpicture} = \frac{C_{Qt}^{(1)}+\left(c_F-\frac{c_A}{2}\right) C_{Qt}^{(8)}}{C_{tG}} \,K_{tG} \times 
            \begin{tikzpicture}[baseline=(4F)]
        \begin{feynman}[small]
            \vertex  (g1)  {$g$};
             \vertex (gtt1) [square dot, scale=\sizesqdot, right= 30 pt of g1, color=gray] {};
            \vertex  (t1) [above right= 25 pt of gtt1] {$t$};
            \vertex (t2) [below right= 25 pt of gtt1] {$t$};

            \diagram* {
                (g1)  -- [gluon] (gtt1),  
                (t2) -- [fermion, line width = 1.5 pt] (gtt1) -- [fermion, line width = 1.5 pt] (t1),};
        \end{feynman}
    \end{tikzpicture}
    \label{eq:diag_tg},
\end{equation}
where we find
\begin{equation}
K_{tG} =
\begin{cases}
 -\frac{\sqrt{2}m_tg_s}{16\pi^2v}& \text{(NDR)}\\
0 & \text{(BMHV)}.
\end{cases}
\label{eq:KtG}
\end{equation}

Since the Lorentz structure of the correction to the gauge vertex 
is similar to the insertion of a chromomagnetic operator, 
the diagrams in Fig.~\ref{fig:feyndiag_4top} acquire a UV divergence
which, analogous to the case of the chromomagnetic operator, can be absorbed by a (now two-loop) counterterm of $C_{HG}$. 
In the $\msbar$ scheme its explicit form is
\begin{equation}
    \begin{split}
       \delta_{C_{HG}}^\text{4-top}=\frac{1}{\eps}\frac{\left(4\pi e^{-\gamma_E}\right)^{2\eps}}{\left(16\pi^2\right)^2}\frac{(-4)g_s^2m_t^2}{v^2}T_F
       \left(C_{Qt}^{(1)}+\left(c_F-\frac{c_A}{2}\right) C_{Qt}^{(8)}\right)\;.
    \end{split}\label{eq:HG-4ferm-ct}
\end{equation}
Schematically, we therefore find
\begin{equation}
\begin{split}
\vcenter{\hbox{\begin{tikzpicture} 
    \begin{feynman}[small]
        \vertex  (g1)  {$g$};
        \vertex  (gtt1) [dot,scale=\sizedot,right =\lentrismall pt  of g1] {};
        \vertex (4F) [dot, scale = \sizesqdot,color=gray,right=\lentrieq/2 pt of gtt1] {};
        \vertex (htt) [dot,scale=\sizedot,below right =\lentrieq pt of 4F] {};
        \vertex  (gtt2) [dot,scale=\sizedot,below left =\lentrieq pt of htt] {};
        \vertex  (g2) [left=\lentrismall+\lentrieq/2 pt of gtt2]  {$g$};
        \vertex (hhh) [dot,scale=\sizedot,right =\lentrieq pt  of htt] {};
        \vertex (h1) [above right =\lenboxeq pt of hhh] {$h$};
        \vertex (h2) [below right =\lenboxeq pt of hhh] {$h$};
        \diagram* {
            (g1)  -- [gluon] (gtt1),
            (g2) -- [gluon] (gtt2),
            (hhh)  -- [scalar] (htt),
            (h1)  -- [scalar] (hhh) -- [scalar] (h2),
            (4F) -- [fermion, line width = 1.5 pt] (htt) 
            -- [fermion, line width = 1.5 pt] (gtt2) -- [fermion, line width = 1.5 pt] (4F), 
             (4F) -- [fermion, line width = 1.5 pt, half left] (gtt1) -- [fermion, line width = 1.5 pt, half left] (4F)
};
    \end{feynman}
\end{tikzpicture}}}
+
\vcenter{\hbox{\begin{tikzpicture} 
    \begin{feynman}[small]
        \vertex  (g1)  {$g$};
        \vertex (hgg) [square dot, scale=\sizesqdot,below right =\lenboxeq pt of g1,color=gray!50] {};
        \node[shape=star,star points=4,star point ratio = 5,fill=black, draw,scale = 0.15, rotate=45] at (hgg) {};
        \node[shape=rectangle,scale = \sizerectangle,color=black,draw] at (hgg) {};
        \vertex  (g2) [below left=\lenboxeq pt of hgg]  {$g$};
        \vertex (hhh) [dot,scale=\sizedot,right =\lentrieq pt  of hgg] {};
        \vertex (h1) [above right =\lenboxeq pt of hhh] {$h$};
        \vertex (h2) [below right =\lenboxeq pt of hhh] {$h$};
        \diagram* {
            (g1)  -- [gluon] (hgg),
            (g2) -- [gluon] (hgg),
            (hhh)  -- [scalar] (hgg),
            (h1)  -- [scalar] (hhh) -- [scalar] (h2),
};
    \end{feynman}
\end{tikzpicture}}}
&=
\frac{C_{Qt}^{(1)}+\left(c_F-\frac{c_A}{2}\right) C_{Qt}^{(8)}}{C_{tG}}\,K_{tG}\left({\cal M}_{tG}^{(a)}+{\cal M}_{tG}^{(b)}\right)
\\
\vcenter{\hbox{\begin{tikzpicture} 
    \begin{feynman}[small]
        \vertex  (g1)  {$g$};
        \vertex  (gtt1) [dot,scale=\sizedot,right =\lentrismall pt  of g1] {};
        \vertex (4F) [dot, scale = \sizesqdot,color=gray,right=\lentrieq/2 pt of gtt1] {};
        \vertex (htt1) [dot,scale=\sizedot,right =\lenboxeq pt of 4F] {};
        \vertex (htt2) [dot,scale=\sizedot,below =\lenboxeq pt of htt1] {};
        \vertex  (gtt2) [dot,scale=\sizedot,left =\lenboxeq pt of htt2] {};
        \vertex  (g2) [left=\lentrismall+\lentrieq/2 pt of gtt2]  {$g$};
        \vertex (h1) [right =\lentrismall+\lentrieq/2  pt of htt1] {$h$};
        \vertex (h2) [right =\lentrismall+\lentrieq/2  pt of htt2] {$h$};
        \diagram* {
            (g1)  -- [gluon] (gtt1),
            (g2) -- [gluon] (gtt2),
            (h1)  -- [scalar] (htt1),
            (htt2) -- [scalar] (h2),
            (4F) -- [fermion, line width = 1.5 pt] (htt1) -- [fermion, line width = 1.5 pt] (htt2) 
            -- [fermion, line width = 1.5 pt] (gtt2) -- [fermion, line width = 1.5 pt] (4F), 
             (4F) -- [fermion, line width = 1.5 pt, half left] (gtt1) -- [fermion, line width = 1.5 pt, half left] (4F)
};
    \end{feynman}
\end{tikzpicture}}}
&=
\frac{C_{Qt}^{(1)}+\left(c_F-\frac{c_A}{2}\right) C_{Qt}^{(8)}}{C_{tG}}\,K_{tG}{\cal M}_{tG}^{(c)}\;,
\end{split}\label{eq:gghh-gauge-4F}
\end{equation}
where ${\cal M}_{tG}^{(a/b/c)}$ denote the amplitude of diagram types (a), (b) and (c)  of Fig.~\ref{fig:feyndiag_chromo}, respectively.

\section{Relations between Wilson coefficients in different $\gamma_5$-schemes}

Scheme dependent contributions such as eq.~(\ref{eq:KtG}) also arise
in the corrections to the top-quark propagator and to the top-Higgs
coupling.
This scheme dependence has the same structure as the one in the
process $gg\to h$ which is described in detail in Ref.~\cite{DiNoi:2023ygk}.
The differences in the NDR and BMHV schemes originating from the mixing of  four-fermion operators with chiral structure $(\bar{L}{L})(\bar{R}{R})$ into the chromomagnetic operator are well known in the context of flavour physics, where is was found that his effect can induce a scheme-dependent anomalous dimension matrix~\cite{Ciuchini:1993fk,Ciuchini:1993ks,Buras:1993xp, Herrlich:1994kh,Dugan:1990df}.
The strategy proposed in \cite{Ciuchini:1993fk,Ciuchini:1993vr, Buras:1993xp} was to perform a finite renormalisation of the chromomagnetic operator to ensure a scheme-independent anomalous dimension matrix.
%
However, when calculating a physical amplitude, 
the scheme dependence involving $C_{Qt}^{(1)}$ and $C_{Qt}^{(8)}$ must be compensated by scheme dependent values for the other parameters of the Lagrangian, 
resulting in an overall scheme independence of the EFT prediction. 
The $\gamma_5$ schemes hence represent equivalent parameterisations of the new physics effects 
and a translation between the two schemes can be achieved by means of finite shifts of the Lagrangian parameters.
The explicit form of the translation relation
between the NDR and the BMHV scheme 
in terms of parameter shifts, derived in Refs.~\cite{DiNoi:2023ygk,Heinrich:2023rsd} is the following (with the top quark mass renormalisation in the on-shell scheme):
\begin{equation}
\begin{aligned}
    \delta{m_t^\text{4-top;\,\text{BMHV}}}&=\delta{m_t^\text{4-top;\,\text{NDR}}}-\frac{m_t^3}{8\pi^2\Lambda^2}\left(C_{Qt}^{(1)}+c_F C_{Qt}^{(8)}\right)
    \\
    C_{tH}^\text{BMHV}&=C_{tH}^\text{NDR}+\frac{\sqrt{2}m_t\left(4m_t^2-m_h^2\right)}{16\pi^2v^3}\left(C_{Qt}^{(1)}+c_F C_{Qt}^{(8)}\right)
    \\
    C_{tG}^\text{BMHV}&=C_{tG}^\text{NDR}-\frac{\sqrt{2}m_tg_s}{16\pi^2v}\left(C_{Qt}^{(1)}+\left(c_F-\frac{c_A}{2}\right)
      C_{Qt}^{(8)}\right)\;.
\end{aligned}\label{eq:scheme_translation}
\end{equation}
Eq.~\eqref{eq:scheme_translation} describes a translation scheme, rather than suggesting
parameter combinations in which the scheme dependence is absorbed, 
as the latter  would require to define a `canonical scheme'.
In order to avoid such an arbitrary choice in physical predictions within SMEFT, combinations of 
Wilson coefficients which allow to cancel the scheme dependence at a given order should be considered.
When matching to concrete models such relations are automatically fulfilled~\cite{DiNoi:2023ygk}.

\section{Phenomenological results}

The  results presented in the following were obtained for a centre-of-mass energy of 
$\sqrt{s}=13.6$\,TeV 
using the PDF4LHC15{\tt\_}nlo{\tt\_}30{\tt\_}pdfas~\cite{Butterworth:2015oua}
parton distribution functions, interfaced to our code via
LHAPDF~\cite{Buckley:2014ana}, along with the corresponding value for
$\alpha_s$.  We used $m_h=125$\,GeV, the top quark mass has been fixed to  $m_t=173$\,GeV to be coherent with the virtual two-loop amplitude calculated numerically.
We set the central renormalisation and factorisation
scales to $\mu_R=\mu_F=m_{hh}/2$ and use 3-point scale variations unless specified otherwise.

\begin{figure}[htb]
\begin{center}
\includegraphics[width=.47\textwidth,page=1]{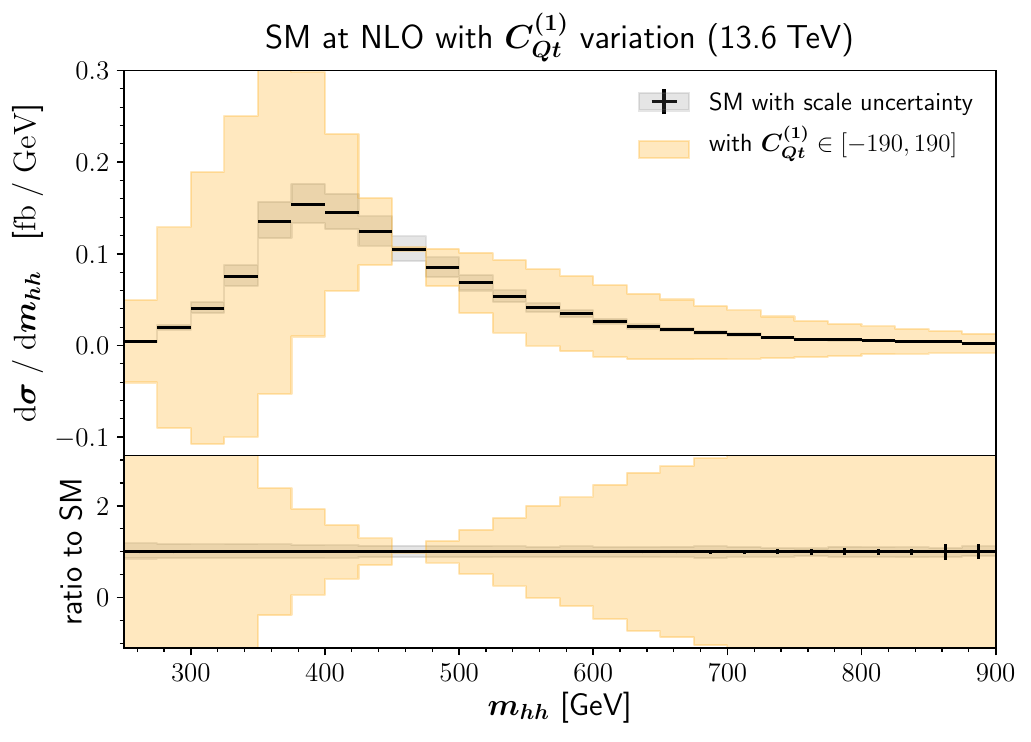}%
\includegraphics[width=.47\textwidth,page=1]{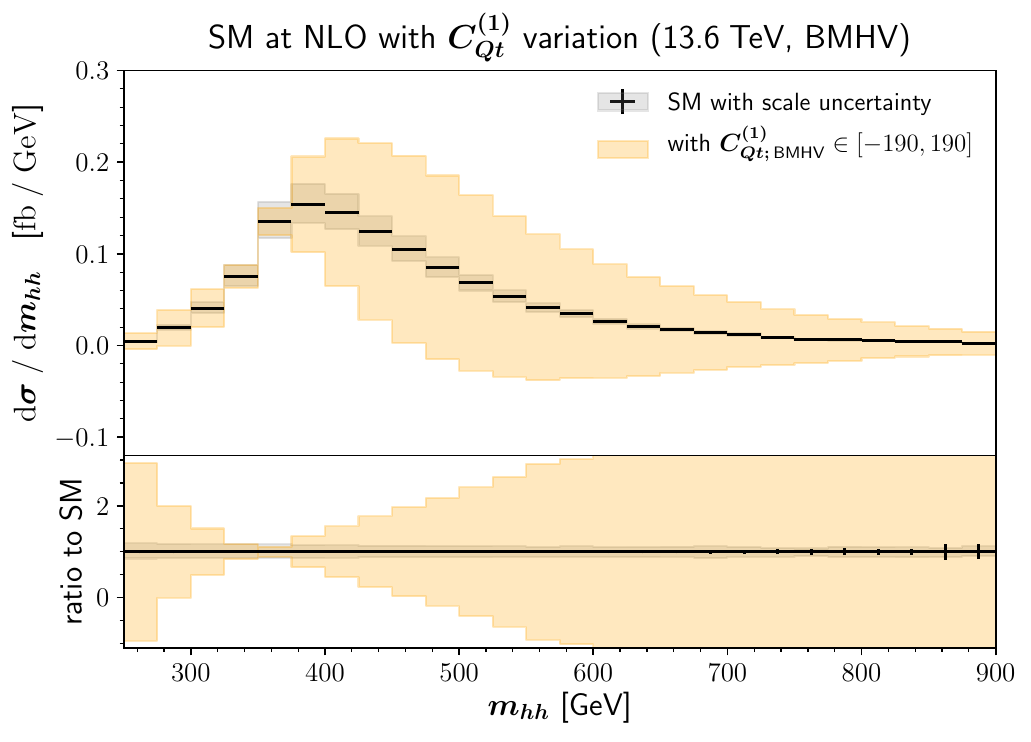}%
    \caption{\label{nlo_CLR_scheme}Effects of $C_{Qt}^{(1)}$-variations on $\mhh$-distributions comparing $\gamma_5$-schemes. 
    Left: NDR scheme, right: BMHV scheme. The interval for  $C_{Qt}^{(1)}$ is oriented at ${\cal O}\left(\Lambda^{-2}\right)$ constraints from Ref.~\cite{Ethier:2021bye}.}
\end{center}
\end{figure}

To demonstrate the effect of different $\gamma_5$-schemes on an individual, scheme dependent Wilson coefficient, we show the Higgs boson pair invariant mass distribution, $\mhh$, where we only include $C_{Qt}^{(1)}$  on top of the SM contribution in Fig.~\ref{nlo_CLR_scheme}. We vary $C_{Qt}^{(1)}$ in the interval $-190\leq C_{Qt}^{(1)}\leq 190$, a range that is inspired by marginalised fits described in Ref.~\cite{Ethier:2021bye}. The grey band denotes the SM scale uncertainties.

Even though the variation range is debatable due to the lack of tight constraints, it is obvious that the scheme differences can be very large for individual Wilson coefficients.
For the case of   $C_{Qt}^{(1)}$, in NDR (left),  the low  $\mhh$-regions exhibits a very large effect way beyond the SM scale uncertainties, 
with unphysical cross sections at very low $\mhh$ values and a sign change around $m_{hh}\sim 460$\,TeV.
This behaviour changes significantly in BMHV (right): there are much weaker effects in the low $m_{hh}$-region, 
the sign change occurs around $m_{hh}\sim 360$\,TeV and the deviation in the high $m_{hh}$-region is more pronounced.

We would like to point out that we have combined these operators with the leading SMEFT operators including NLO QCD corrections as described in Refs.~\cite{Heinrich:2022idm,Heinrich:2023rsd}.
This combination is provided as an extension to the public {\tt ggHH\_SMEFT} code as
part of the {\tt POWHEG-Box-V2}~\cite{Alioli:2010xd}.

\section{Conclusions}

We have discussed the calculation of contributions from the chromomagnetic operator and 4-top operators 
to Higgs boson pair production in gluon fusion and argued that
these operators both appear at the same  order in a power counting scheme that takes into account whether  dimension-6 SMEFT operators are loop-generated or (potentially) tree-generated.
We have shown that the contributions of those Wilson coefficients, when considered individually, depend on the chosen $\gamma_5$-scheme, and we have provided relations that allow a translation between the NDR and BMHV schemes. The explicit example of the 4-top-operator $C_{Qt}^{(1)}$ illustrates 
that the differences induced by a scheme change can be larger than the SM scale uncertainties.
To obtain meaningful results for constraints on such Wilson coefficients, it is therefore recommended not to study  those  coefficients which are connected through scheme translations in isolation,
as only their combination is a scheme independent parametrisation of BSM physics at the considered order in the power counting.

\section*{Acknowledgements}

We would like to thank Stephen Jones and Matthias Kerner for
collaboration related to the $ggHH$@NLO
projects and Gerhard Buchalla for useful discussions.
This research was supported by the Deutsche Forschungsgemeinschaft (DFG, German Research Foundation) under grant 396021762 - TRR 257.



\bibliographystyle{JHEP}
\bibliography{LL2024_SMEFT.bib}

\end{document}